\documentclass[fleqn,usenatbib]{mnras}

\usepackage{newtxtext,newtxmath}

\usepackage[T1]{fontenc}

\DeclareRobustCommand{\VAN}[3]{#2}
\let\VANthebibliography\thebibliography
\def\thebibliography{\DeclareRobustCommand{\VAN}[3]{##3}\VANthebibliography}

\usepackage{graphicx}	
\usepackage{amsmath}	
\usepackage[dvipsnames]{xcolor}

\title[MaNGA constraints on modified gravity]{Testing screened modified gravity with SDSS-IV-MaNGA}

\author[R. G. Landim et al.]{
Ricardo G. Landim,$^{1}$\thanks{E-mail: ricardo.landim@port.ac.uk}
Harry Desmond,$^{1}$\thanks{E-mail: harry.desmond@port.ac.uk}
Kazuya Koyama,$^{1}$
Samantha Penny$^{1}$
\\
$^{1}$Institute of Cosmology and Gravitation, University of Portsmouth, Dennis Sciama Building, Portsmouth PO1 3FX, United Kingdom\\
}

\date{Accepted XXX. Received YYY; in original form ZZZ}

\pubyear{2015}

\begin{document}
\label{firstpage}
\pagerange{\pageref{firstpage}--\pageref{lastpage}}
\maketitle

\begin{abstract}
Fifth forces are ubiquitous in modified gravity theories, and must be screened to evade stringent local tests. This can introduce unusual behaviour in galaxy phenomenology by affecting galaxies' components differently. Here we use the SDSS-IV-MaNGA dataset to search for a systematic excess of gas circular velocity over stellar circular velocity, expected in thin-shell-screened theories in the partially screened regime. Accounting for asymmetric drift and calibrating our model on screened subsamples, we find no significant evidence for a screened fifth force. We bound the fifth-force strength to $\Delta G/G_\text{N} < 0.1$ for all astrophysical ranges, strengthening to $\sim$0.01 at Compton wavelength of 3 Mpc for the Hu-Sawicki model, for instance. This implies a stringent constraint on scalar--tensor theories, for example $f_{\mathcal{R}0} \lesssim 10^{-8}$ in Hu--Sawicki $f(\mathcal{R})$ gravity.
\end{abstract}




\section{Introduction}

Dark energy is one of the greatest mysteries in modern cosmology, responsible for the current phase of accelerated expansion of our Universe \citep{1998AJ....116.1009R, 1999ApJ...517..565P}. The simplest candidate is a cosmological constant, however the large discrepancy between the observed value and a theoretical prediction causes an issue that is long known \citep{1989RvMP...61....1W}. Due to the lack of satisfactory theoretical explanation for the cosmological constant and current tensions between datasets \citep[e.g.][]{2021CQGra..38o3001D, 2022JHEAp..34...49A}, alternatives to the standard $\Lambda$-Cold-Dark-Matter ($\Lambda$CDM) paradigm have been investigated. Among a plethora of candidates, modified gravity (MG) comprehends a broad class of models that changes the standard setup of General Relativity. 

Most viable MG models implement a \emph{screening mechanism}, where the fifth force that arises from extra gravitational degrees of freedom is suppressed in high density regions. This renders MG effects in the Solar System negligible, where precise constraints on the inverse square law and strength of gravity have been set (for recent reviews, see~\citealt{2018LRR....21....1B,2019arXiv190803430B,Brax,2024arXiv240514638F}).
Astrophysics can constrain screened MG models, through stellar evolution as well as the structure and dynamics of galaxies \citep{2018LRR....21....1B, 2019arXiv190803430B, Brax}.  In this last category of tests lies the rotation curves of galaxies,  which can be applied to a class of screened  models, where the force inside the so-called screening radius ($r_s$) of a spherical object, for instance,  is simply Newtonian, whereas outside the screening radius  a fifth force is non-negligible. Outside the object (with radius $R$),  fifth force is sourced by the mass within a thin-shell of thickness $R-r_s$.  Candidates in this thin-shell classification are, for instance, chameleon \citep{2004PhRvD..69d4026K}  symmetron \citep{2010PhRvL.104w1301H} and environment-dependent dilaton \citep{2010PhRvD..82f3519B}.  Some forms of $f(\mathcal{R})$ theories, such as the Hu-Sawicki (HS) model \citep{2007PhRvD..76f4004H}, present a chameleon mechanism as well, as an example of equivalent formalism between $f(\mathcal{R})$ and scalar-tensor theories \citep{2008PhRvD..78j4021B}. The fifth force actuates along with the Newtonian gravity for the (low-density) gases in the galaxy (thus increasing the gravitational constant), whereas the dense main-sequence stars are self-screened and hence do not have their rotation curves changed (a violation of the weak equivalence principle). Due to this effect, it is expected that gas rotation curves are enhanced at least a few percent when compared to the stellar rotation curves. In the HS model, for example, the Newton constant is increased by 1/3, leading to an increase of roughly $\sqrt{1/3}$ in the gas rotation curve. Studies using rotation curves to constrain these MG models were pioneered by \citet{2009PhRvD..80j4002H,2011JCAP...10..032J} and are still few \citep{2013JCAP...08..020V,2014arXiv1407.6044V}, mainly due to the low number of data available until recently. 

The situation changed in the past years, when the Sloan Digital Sky Survey (SDSS) programme Mapping Nearby Galaxies at Apache Point Observatory \citep[MaNGA;][]{2015ApJ...798....7B, 2015AJ....149...77D,2016AJ....151....8Y,2015AJ....150...19L,2016AJ....152...83L,2016AJ....152..197Y,2017AJ....154...86W,2019AJ....158..231W,2019AJ....158..160B,2021AJ....161...52L} obtained the dynamics of more than 10,000 galaxies. Not only were the 2-D velocity field maps measured for the target galaxies (from which rotation curves can be obtained), but also the velocity dispersion maps.  
The measured velocity dispersion of stars is a projection of the phase-space distribution of their orbits along the line-of-sight.  I.e. at any given position, the stars observed within an aperture are in different phases of their elliptical orbits, which leads to a variance in their line-of-sight velocities.  The connection between the phase-space distribution function and asymmetric drift (AD) can be derived from moments of the collisionless Boltzmann equation.  It can be thought of as an artifact of the radial decrease in both the stellar density and the stellar velocity dispersion.
This effect is further explained in Section \ref{sec:ad} and could not be modelled explicitly in previous analyses. 

Therefore, in this work we investigate screened MG using data from SDSS-IV-MaNGA. We focus on thin-shell-screened theories, with HS $f(\mathcal{R})$ as the paradigmatic example, comparing explicitly to the state-of-the-art galaxy morphology constraints of \citet{2020PhRvD.102j4060D}. After extracting the  rotation curves for  gas and stars and calculating the uncertainties on them, we use a Markov Chain Monte Carlo (MCMC) algorithm to obtain upper limits on the variation of the Newton constant ($\Delta G/G_\text{N}$). Our main results are shown in Figures \ref{fig:ad} and \ref{fig:env1only}.

Our paper is structured as follows. Section \ref{sec:mg} contains a quick review on screened MG, whereas Section \ref{sec:dataset} presents an overview of the dataset we use. Section \ref{sec:eq} details the methodology we  developed in order to extract and compare the circular velocities from the MaNGA dataset and Section \ref{sec:results} presents our results. Section \ref{sec:conclusions} is reserved for conclusions.

\section{Screened modified gravity}\label{sec:mg}

We will test the models of MG that present a thin-shell screening mechanism, where the fifth force is negligible in bodies near a dense environment, inside a given radius (the screening radius $r_s$), and non-negligible in a thin-shell outside the screening radius. 

 The chameleon mechanism, for example, is described by an extra scalar field $\varphi$ interacting with  matter fields through a conformal coupling, which produces an effective potential $V_{\rm eff}(\varphi) = \Lambda^{4 + \gamma}\varphi^{-\gamma} +8\pi \alpha_c G_\text{N} \varphi \rho $, where $\alpha_c$ is the coupling, $\rho $ is the mass density, and $\gamma$ is a constant.  The resulting fifth force acceleration is then \citep{2011ApJ...732...25C,2012PhRvD..85l3006D,2019arXiv190803430B}:
 \begin{align}
     a_5 &= 2\alpha_c^2 \frac{G_N M(r)}{r^2}\left[1- \frac{M(r_s)}{M(r)}\right] \qquad r_s< r< R_{\max} \\
     a_5 & = 2\alpha_c^2 \frac{G_N M }{r^2}\left[1- \frac{M(r_s)}{M}\right]e^{-m_{\rm eff }(\varphi_0) r} \quad   r> R_{\max}\,,  
 \end{align}
 where $m_{\rm eff }= \gamma(\gamma + 1)\Lambda^{\gamma +4}\left(\frac{\alpha_c\rho}{n M_{\rm Pl}\Lambda^{\gamma+4}}\right)^{\frac{ \gamma+2}{\gamma +1}}$ is the effective mass, $R_{\max} $ is the radius of the object and $\varphi_0$ is the background field value (where the total field is expanded around the background field sourced by the surrounding environment $\varphi \rightarrow \varphi_0 + \varphi$).
 
Effectively, the total acceleration is now the sum of the gravitational one and the fifth force $a_5$, resulting in an increase of the Newton constant $(1+2\alpha_c^2)G_\text{N}$ per unit mass for unscreened objects. A large population of stars are expected to be self-screened so that they would not feel the fifth force \citep{2012PhRvD..85l3006D}. The gas component, on the other hand, is not as dense as stars and it would have an enhancement to its circular velocity due to the extra force.  Therefore, the ratio of the gas circular velocity to the stellar circular velocity will show whether there is a screening mechanism of this kind:
\begin{equation}
    \frac{V^2_{c,g}}{V^2_{c,*}}=1+2\alpha_c^2 = 1+\frac{\Delta G}{G_\text{N}}\,.
\end{equation}

An example of $f(\mathcal{R})$ theory that presents the screening mechanism is the HS model \citep{2007PhRvD..76f4004H} 
\begin{equation}
    f(\mathcal{R}) = -m^2\frac{c_1(\mathcal{R}/m^2)^n}{1+c_2(\mathcal{R}/m^2)^2}\,,
\end{equation}
where $\mathcal{R}$ is the Ricci scalar, $m^2=H_0^2\Omega_m$ at $z=0$, $c_1$, $c_2$ and $n$ are free parameters, $H_0$ is the Hubble rate today and $\Omega_m$ is the density parameter for non-relativistic matter. Here we follow previous  literature and set $n=1$. One of the two remaining free parameters is fixed by matching the $\Lambda$CDM expansion history to first order, whereas the second parameter is related to the derivative of $f$ with respect to $\mathcal{R}$ evaluated today $f_{\mathcal{R}0}$. This ends up being the only free parameter and in this model it is related with the Compton wavelength $\lambda_C \equiv \mu^{-1}$, where $\mu$ is the  scalar field mass. Therefore, only regions inside $\lambda_C$ feel the fifth force\footnote{We assume that the mass of the $f(\mathcal{R})$ scalar does not change inside a galaxy. However, it might instead be that it varies, affecting, therefore, the Compton wavelength.  In this situation, the Compton wavelength may become shorter than the interstellar separation, meaning that the scalar field can resolve individual stars, rather than being affected globally, in the standard scenario, by  the stellar component of the disc as a smoothly varying density field. This would result in a, albeit small, shift in where the screening radius is (C. Burrage, B. March and A. Naik, private communication). }
and one can write $\lambda_C$ in the HS model as \citep{2007PhRvD..76f4004H, 2012JCAP...07..034C}
\begin{equation}
    \lambda_C \simeq 0.32 \sqrt{f_{\mathcal{R}0}/10^{-8}}\, \text{Mpc}\,.
\end{equation}
 We will use this relation to reduce the dimensionality of the inference parameter space, but when including environmental screening, we will relax this relation between $\lambda_C$ and $f_{\mathcal{R}0}$.  

Thin-shell-screened MG has been tested in various ways \citep{Brax}. Laboratory tests using atom interferometry have put stringent constraints on  chameleon and symmetron parameter spaces \citep[][see fig. 1 and 2 of \citealt{Brax}]{2017NatPh..13..938J, 2019PhRvL.123f1102S}. Regarding astrophysical tests, screened models are often presented through the $n = 1$ HS model, which is constrained through  comparisons between Cepheid and Tip of the Red Giant Branch distance indicators \citep{2013ApJ...779...39J,2019PhRvD.100d3537D,2023PhRvD.108l4050H, 2023PhRvD.108b4007H}, the rotation curves of late-type galaxies \citep{2018MNRAS.480.5211N,2019MNRAS.489..771N} and the rotation curves of dwarf galaxies \citep{2014arXiv1407.6044V}.
The most recent and powerful analysis of the HS model investigated the warps and gas-star offsets  of thousand galaxies observed in SDSS and ALFALFA, excluding  $f_{\mathcal{R}0}>1.4 \times 10^{-8} $  at  $ 1\sigma$ confidence within $0.3\leq \lambda_C \leq 8$ Mpc \citep{2018PhRvD..98f4015D, Desmond_reconstructing,2018PhRvD..98h3010D,2020PhRvD.102j4060D}. 

\section{Data set}\label{sec:dataset}

MaNGA is an SDSS-IV survey \citep{2015ApJ...798....7B, 2015AJ....149...77D,2016AJ....151....8Y,2015AJ....150...19L,2016AJ....152...83L,2016AJ....152..197Y,2017AJ....154...86W,2019AJ....158..231W,2019AJ....158..160B,2021AJ....161...52L} that functioned between 2014 and 2020 
 and obtained spectra for over 10,000 galaxies with 
$M_* > 10^9 M_{\odot}$, across $\sim  2700$ deg$^2$ and redshift rage $0.01 \leq z\leq 0.15$, using 17 integral field units,  with a wavelength coverage of 3500 -- 10000 \r{A} and spectral
resolution $R\sim  2000$. Our dataset contains the most recent MaNGA catalog \citep[DR-17]{2022ApJS..259...35A},\footnote{Retrieved  using scripts described in \url{https://github.com/ricardoclandim/NIRVANA} or \url{https://github.com/kbwestfall/NIRVANA}.} in which we use MaNGA's hybrid binning ({\tt HYB10}) with the {\tt MILESHC} template  to determine the stellar kinematics and the MaStar stellar library ({\tt MASTARSSP}) to fit the (gas) emission lines. MaNGA produced the velocity maps of several emissions lines, but all the velocities are tied together.  So the velocity measurements are effectively a single velocity fit to all emission lines.  The velocity dispersions are specific to each emission line, and here we focus our analysis on  H$\alpha$. We tested different templates and the results for our purposes are near-identical. 

Kinematic data for our sample was provided by the MaNGA Data Analysis Pipeline \citep[DAP;][]{2019AJ....158..231W, 2019AJ....158..160B}. Briefly, the MaNGA DAP provides high-level spatially-resolved data products for all galaxies in MaNGA including stellar kinematics (velocity and velocity dispersion) and emission line properties such as flux, equivalent widths and kinematics. In particular, emission line kinematics for several ionisation species are included, making the DAP output ideal for studies of galaxy kinematics. For further information on how the galaxy kinematics are measured, see \citet{2019AJ....158..231W}. 

Given our study requires both stellar and gas kinematics, we restrict our sample to objects with blue optical colours $u-r < 2.0$ and galaxy mass $M_{\star} > 5\times10^{9}\,\mathrm{M_{\odot}}$. The colour-cut removes redder galaxies in which the gas component may be truncated or absent though evolutionary or environmental processes, i.e., galaxies residing on the red-sequence of the colour-magnitude relation \citep[e.g.][]{2014MNRAS.440..889S}. The mass cut avoids the inclusion of dwarf or low-mass galaxies in which the kinematics are more susceptible to disruption by galaxy-galaxy interactions.\footnote{Low-mass galaxies are also those which would show the largest signal for our analysis, as they are most likely to be unscreened. This cut therefore renders our results conservative as well as removing a potential systematic, as we are forced to infer fifth-force parameters from less optimal systems.} These cuts resulted in a sample of 2220 galaxies with redshift $0.007< z< 0.150$ and median S\'ersic index $n=1.80$, showing our sample is disc-galaxy dominated.

\section{Methodology }\label{sec:eq}

In this section we describe the methodology we developed and used to produce the rotation curves and velocity dispersion profiles and to analyse the data, in order to obtain the results shown in the next section.

We fit the MaNGA velocity fields using code within the Nonaxisymmetric Irregular Rotational Velocity ANAlysis ({\tt NIRVANA}; DiGiorgio Zanger et al., submitted) framework.  The primary goal of the software is to model bisymmetric features in galaxy velocity fields; however, it also provides modules for performing more simple, axisymmetric fits that model the gas and stellar kinematics simultaneously.  Additionally, {\tt NIRVANA} performs sophisticated beam-smearing corrections that account for observational effects on the measured rotation curves and velocity dispersion profiles.  Relevant to our study here, we use {\tt NIRVANA} to optimally model the bulk recession velocity ($V_{\rm sys}$), coordinates of the dynamical center ($x_0,y_0$), the disk inclination ($i$), and the on-sky position angle (PA; $\phi_0$).  Uncertainties in these geometric projection parameters are propagated through our calculation of the in-plane rotation speeds for both the gas and stellar kinematic traces.

Parameter uncertainties are calculated using the Hessian matrix computed as part of the Levenberg-Marquardt non-linear least-squares optimization procedure; see Section 15.5 of \citet{nr}.  This approach assumes the $\chi^2$ space is smooth and unimodal, that the kinematics have independent Gaussian errors, and that $\chi_\nu^2 \sim 1$.  In particular, when the latter is not true, errors calculated following this approach can be over- or under-estimated.  We mitigate these inaccuracies in the parameter uncertainties by introducing ``intrinsic scatter'' terms for both the velocity and velocity dispersion.  These terms are added in quadrature with the measurement errors and iteratively optimized during the fitting procedure to yield an error distribution about the best-fitting model that is nearly Gaussian such that $\chi^2_\nu \sim 1$.  This improves our model parameter uncertainties, modulates the sensitivity of the modeling to underestimated kinematic errors, and provides a useful metric for the significance of deviations about the best-fit models.

\subsection{Model for rotation and dispersion curves}\label{sec:rotcurvtheory}

\texttt{NIRVANA} enables us to model galaxy rotation curves and velocity dispersion profiles using simple analytic forms for the rotation curves and velocity dispersion profiles.  For this study, we choose:
\begin{align}
    V_{\rm rot, m}(R)&= V_0\tanh\left( \frac{R}{R_0}\right)\,,\label{eq:vrot_tanh}\\
    \sigma_{R,m}(R)&=\sigma_0e^{-R/h}\,,\label{eq:vdisp_exp}
\end{align}
where $V_0$,  $\sigma_0$, $R_0$, $h$ are free parameters that are optimized simultaneously with the five parameters given above ($V_{\rm sys}, x_0, y_0, i, \phi_0$).  Both of these functional forms are chosen as approximations for the well-known empirical shape exhibited by galaxies: the hyperbolic tangent rises smoothly to an asymptotically flat value, as empirically expected of observed rotation curves, and its form is close to a pseudo-isothermal sphere \citep[e.g.][]{2001AJ....122.2396D}.  For the velocity dispersion, galaxy disks have long been known to have exponential (S\'ersic $n\approx 1$) surface brightness profiles \citep{1970ApJ...160..811F}, which can be tied directly to their surface mass density distributions \citep[e.g.][]{1988A&A...192..117V} and shown to be a natural consequence of the galaxy formation process \citep[e.g.][]{2009MNRAS.396..121D}.  Assuming a constant mass-to-light ratio in the disk (correct to first order), it is expected that the stellar velocity dispersion also follows an exponential form \citep[e.g.][]{1988A&A...192..117V} as demonstrated by observations \citep[e.g.][]{2013A&A...557A.130M}. 

We emphasise that the choice of these functional forms, and indeed the modeling of the galaxy kinematics using \texttt{NIRVANA}, is primarily to determine the best-fitting on-sky projection of each galaxy disk.  Nevertheless, we investigate possible systematic effects due to our choice of these specific functions by also assuming modified forms in Appendix \ref{sec:appendix}.  This alternate model should not be strictly taken as an equivalent substitute of our fiducial model because the latter generally fits better than the former. In our case, the second model fits 991 galaxies with a reduced $\chi^2$ between 0.95 and 1.05, whereas the fiducial one fits 1110 with the same reduced $\chi^2$ range, i.e. 119 more galaxies than the alternative model.

\subsection{Asymmetric drift correction}\label{sec:ad}

The AD equation is derived by taking the $v_R$ moment of the collisionless Boltzmann equation \citep{2008gady.book.....B}:
\begin{equation}\label{eq:ad_core}
V^2_{c} = V^2_{\rm rot} - \sigma^2_R\left(\frac{d \ln \sigma^2_R}{d \ln R} + \frac{d \ln \rho_V}{d \ln R} \right)\,,
\end{equation}
where $V_c$ is the circular speed defined by the gravitational potential and $V_{\rm rot}$, $\sigma_R$, and $\rho_V$ are, respectively, the average tangential speed, radial component (in a cylindrical coordinate frame with the disk midplane at $z=0$) of the velocity dispersion, and volume density of a dynamical tracer.  Assuming the disk has a constant scale height, one can write $\rho_V\propto \Sigma_V$, where  $ \Sigma_V $ is the  surface density.  Furthermore, for a locally isothermal disk, the stellar disk-mass surface density is proportional to the vertical (out-of-plane) component of the disk stellar velocity dispersion $\sigma_z$, $\Sigma_V \propto \sigma_z^2$ \citep{2010ApJ...716..198B}. If the axial ratio $\alpha =  \sigma_z/\sigma_R$ is independent of the radius and we take an exponential profile for the velocity dispersion ($\sigma_R \propto \sigma_0e^{-R/h}$), Eq. (\ref{eq:ad_core}) becomes 
\begin{equation}\label{eq:ad_final}
V^2_{c} = V^2_{\rm rot} + \frac{4R}{h}\sigma^2_R\,,
\end{equation}
for the stellar case. For the gas component the relations above do not necessarily hold, but we can assume that the AD correction is similar. In order to investigate a potential uncertainty on this calculation we assume a second hypothetical situation where the gas surface density has a very strong dependence on $\sigma_z$, yielding the  AD correction of $8R/h$, such that the gas circular velocity is larger than the fiducial case (with $4R/h$). This alternative case is illustrative and only aimed to investigate the influence of this correction on the results, but, as we will show, we will restrict the  other analyses to the fiducial case, i.e. with $4R/h$. 

The velocity dispersion of the ionised gas is typically $\sim20$ km/s \citep{2022ApJ...928...58L} and is rather constant as a function of radius (i.e. $h$ for the gas is typically very large compared to the stars).  Contrarily, the velocity dispersion of the stars decreases much more quickly, but can have a maximum value many times larger than the gas velocity dispersion.  The result is that ionised gas rotation curves are much closer to the theoretically defined circular speed than are the stars, such that ionised gas rotation curves are observed to show faster rotation than observed for stars.  The AD corrections are therefore critical for unbiased inference of MG on our rotation curves because AD and MG lead to similar effects on the measured difference between the two tracers (the unscreened gas and the screened stars).

\subsection{Screening radius}\label{sec:scr_radius}

Previous works have usually employed a binary classification when investigating the effects due to the screening  in galaxies, i.e. one assumes that a whole galaxy is either screened or unscreened. Although good as a first approximation, it is not the most robust way to investigate the effect of fifth forces in those objects. A galaxy may be partially screened below a certain ``screening radius'' $r_s$, which may have an important impact on the effect of a fifth force \citep{2023arXiv231019955B}. We assume however that the dark matter halo is spherically symmetric, while in reality it is triaxial; this has been shown to have a small effect on the resultant fifth force~\citep{ellip_1, ellip_2}.

As it was aforementioned, in order to infer the galaxy parameters, we  use a thin-disk model from {\tt NIRVANA} to fit the 2-D velocity field maps. The geometric projection parameters are estimated through the thin-disk model and used to de-project the velocity field and measure rotation velocities and velocity dispersions as a function of radius.
Using the measured circular velocity (after correcting for AD), one can find the screening radius through the relation \citep{2012PhRvD..85l3006D,Sakstein:2013pda,2016JCAP...11..045B, 2018LRR....21....1B}:\footnote{Strictly speaking, this relation only holds for spherically symmetric densities. Here, however, we have an axisymmetric distribution within the MaNGA's data range. The axisymmetry would introduce a $\ln r $ in the integrand, which would reduce a little the integrated value, although not significantly. Use of Eq.~(\ref{eq:chi}) is therefore conservative.}
\begin{equation}\label{eq:chi}
    \chi_c \equiv \frac{\varphi_0}{2\alpha_c M_{Pl}}=\frac{4\pi G}{c^2} \int_{r_s}^{R_{\max}}dr\,r\rho(r)\,,
\end{equation}
where $\varphi_0$ is the background field value and $R_{\max}$ is a maximum cutoff radius. The mass density $\rho(r)$ is numerically calculated from the Poisson equation $\nabla^2\phi= 4\pi G \rho/c^2$ in cylindrical coordinates, and the gravitational potential $\phi$  for the self-screening is axisymmetric but independent of the axial coordinate $z$. The potential, in turn,  is related to the intrinsic circular velocity through $V^2_{c, \rm in}=R \partial \phi/\partial R$. The intrinsic stellar circular velocity is calculated taking one of the two different models into account (Eqs. (\ref{eq:vrot_tanh}) and (\ref{eq:vdisp_exp}) or (\ref{eq:vrot_poly}) and (\ref{eq:vdisp_expbase})), with the model parameters fitted by {\tt NIRVANA} in Eq. (\ref{eq:ad_final}).

In theory $R_{\max}$ should be infinity, or in practice very large, such that the integral should be calculated for larger values than the ones measured by MaNGA. We therefore extrapolate the data by assuming an isothermal sphere  profile  $\rho_{\rm iso} \propto R^{-2}$, but with a maximum radius $R_{\max} = 50 R_{\rm eff}$.\footnote{Other cutoffs were investigated and give near-identical results.} The constant of proportionality is set taking the inferred $\rho$ from each galaxy, as explained in the previous paragraph, at the maximum measured radius. The isothermal sphere profile is only valid outside the range of the data (for the extrapolated radius) and this choice of dependence on $R$ guarantees that the enclosed total mass  of a given galaxy is within the order of magnitude it should be when we compute it through $M= V^2_{\rm asymp} R_{\max}/G$, where $V^2_{\rm asymp}$ is the asymptotic circular velocity.  


In the HS model the self-screening parameter is related to $f_{\mathcal{R}0}$, and in turn to $\lambda_C$ through \citep{2007PhRvD..76f4004H,2012JCAP...07..034C}
\begin{equation}
    \chi_c = \frac{3}{2}f_{\mathcal{R}0}= \frac{3}{2}\times 10^{-8} \left( \frac{\lambda_C}{0.32 \,\text{Mpc}}
 \right)^2\,,
\end{equation}
implying that once we fix one MG parameter, all others are fixed as well. In order to investigate the influence of the environmental screening on the results, we will vary $\chi_c$ in the HS model by raising or lowering it by a factor of 10 from its fiducial dependence for a given $\lambda_C$. This will give an indication of the effect of deviations from the HS $\lambda_C-\chi_c$ relation, although of course larger deviations than this are possible.

\subsection{Environmental screening}

There is an extra contribution for the screening of a given galaxy coming from neighbouring galaxies. This environmental screening contribution comes from the dimensionless Newtonian potential $\Phi_{\rm ext}$ sourced by masses within $\lambda_C$, which in turn will yield a constant term in the right-hand side of Eq. (\ref{eq:chi}). This extra potential is evaluated at $i$ test particles for $j$ point masses via \citep{2011PhRvD..83d4007Z,2011PhRvL.107g1303Z,2012JCAP...07..034C}
 \begin{equation}
     \Phi_{\rm ext,i}(\lambda_C)= \sum_{r_{i,j}<\lambda_C} \frac{G_\text{N} M_j}{c^2 r_{i,j}}\,.
 \end{equation}

We use the method of~\citet{Desmond_reconstructing}, as upgraded in~\citet{2020PhRvD.102j4060D},
to calculate the environmental screening for galaxies up to 250 Mpc, which is the maximum distance in which the  density field reconstructed by {\tt BORG} algorithm \citep{2012MNRAS.425.1042J, 2013MNRAS.432..894J} using the 2M++ survey is reliable \citep{2011MNRAS.416.2840L,2016MNRAS.455.3169L}.

\subsection{Data analysis}

\begin{figure}
    \centering   \includegraphics[width=\columnwidth]{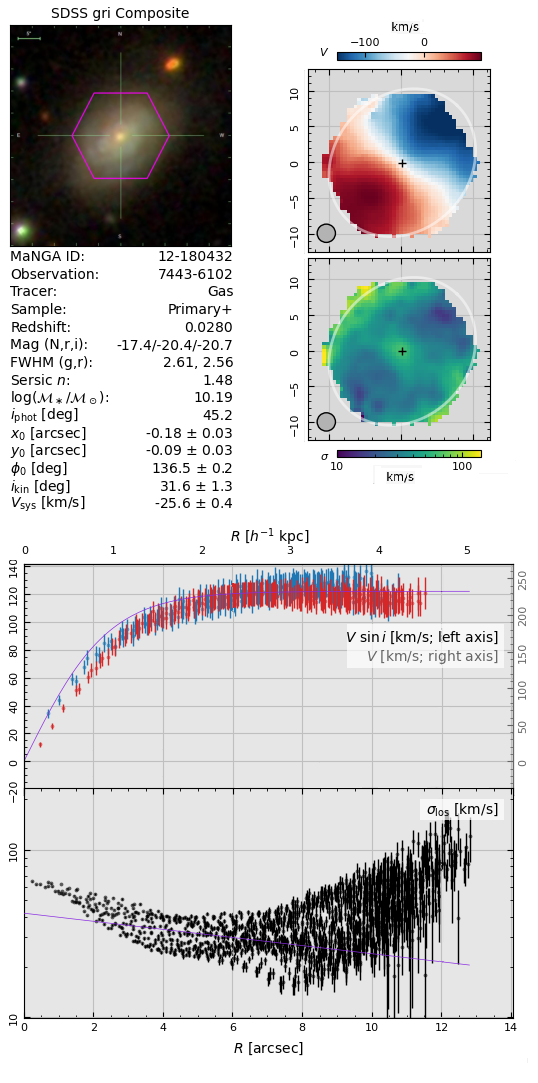}
    \caption{Example of ionised-gas 2-D velocity field and 2-D H$\alpha$ velocity dispersion  field (top right panels). The middle/bottom panels are the gaseous rotation and dispersion curves from {\tt NIRVANA}. The solid violet lines are the theoretical models, blue and red curves correspond to the approaching and receding sides of the galaxy, respectively. 
  The increase at large radii of the dispersion curve is likely due to low signal-to-noise ratio (SNR) and relatively low spectral resolution.
    \protect\footnotemark}
    \label{fig:nirvana}
\end{figure}

We calculate the corresponding errors on the measurements through the combination of the propagation of uncertainties on the galaxy parameters estimated by {\tt NIRVANA} (taking also the covariance of the parameters into account) and the uncertainties on the MaNGA measurements themselves.\footnote{MaNGA has significant covariance between spatially adjacent spaxels, however it is assumed for simplicity that there is no covariance between the spaxels of the same galaxy. Later on, we analyse the case with a fictitious strong covariance between the data points, but the results are more stringent, so we keep the more conservative scenario with only diagonal terms in the covariance matrix.} We modified {\tt NIRVANA} to include the propagation of uncertainties and an example of this version of {\tt NIRVANA}'s output is shown in Figure \ref{fig:nirvana}. 

\footnotetext{We have not imposed additional cuts on the data due to low SNR. Usually low SNR regions have larger uncertainties, but we tested the analysis for large SNR and the results are near-identical.}

Once the rotation curve and dispersion curve in {\tt NIRVANA} are obtained we averaged for each galaxy and tracer the radii ($R/R_{\rm eff}$, where $R_{\rm eff}$ is the effective radius of a given galaxy) and velocities within a bin of $\Delta R/R_{\rm eff} = 0.1$. We found that this value for the bin is sufficient to have detailed velocity curves. After this procedure is done for the two tracers of a galaxy (gas and stars), for the rotation and dispersion curves, the total circular velocity for each tracer is calculated according to Eq. (\ref{eq:ad_final}). The step described in this paragraph is necessary because the measured radii of the 
 (projected) rotation velocity and velocity dispersion  measurements are not exactly the same, neither for stars or gas. The corresponding uncertainty of each averaged point was also calculated. In order to test this step, we interpolated the dispersion and rotation velocity curves and we found that the results are virtually the same. The problem with the use of interpolation instead of this binning in the first place is that the propagation of uncertainties for the interpolated points is more difficult.

\begin{figure}
    \centering
\includegraphics[width=\columnwidth]{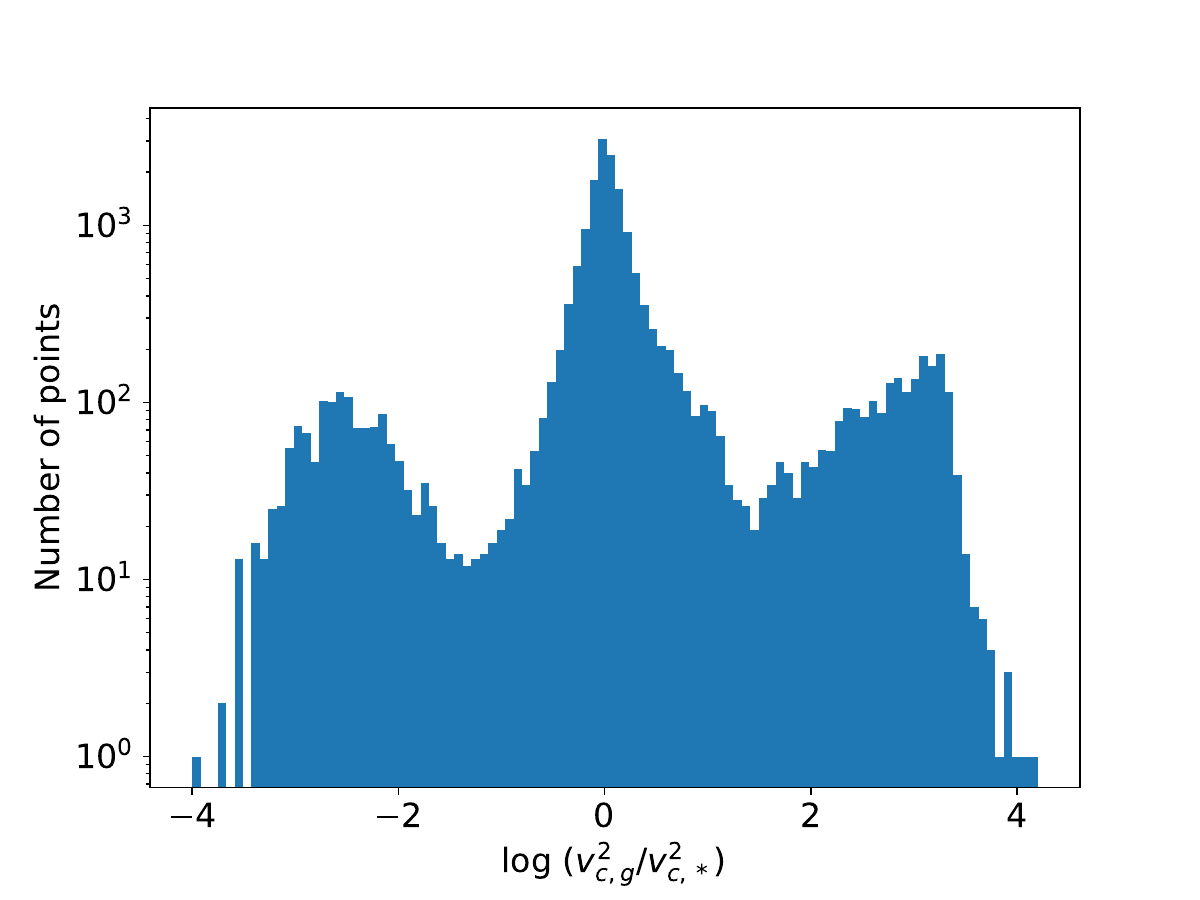}
    \caption{Histogram of data of ratio of circular gas velocity to star circular velocity. The weighted average of these points (where the weights are the inverse of the variances) gives $\langle V^2_{c,g}/V^2_{c, *}\rangle \approx  0.69 $ and the median is 1.06. 
    }
    \label{fig:hist}
\end{figure}

 In order to exclude galaxies that are poorly constrained by {\tt NIRVANA} (due to bars, low-inclination, poor quality data or poor fits of Eqs.~\ref{eq:vrot_tanh} and~\ref{eq:vdisp_exp}) we only keep galaxies whose final estimate in {\tt NIRVANA} had a reduced $\chi^2$ between 0.95 and 1.05. We also check if the five inferred geometric projection parameters for  each galaxy agree between stars and gas. The final sample has 1110 galaxies (with a total of 18,204 points). Using the screening radius described in Section \ref{sec:scr_radius}, we classify which part of a given galaxy is screened or unscreened, for a given $\chi_c$. Then, we select the corresponding data points of $V_{c,g}^2 / V_{c,*}^2$  in the unscreened (or screened) regions of the galaxies, and after that we put the data points of all galaxies together in the same sub-sample, thus creating screened and unscreened subsamples. A histogram of all data points of $V_{c,g}^2 / V_{c,*}^2$ (without dividing them by screening) is shown in Figure \ref{fig:hist}.

\begin{figure}
    \centering
\includegraphics[width=\columnwidth]{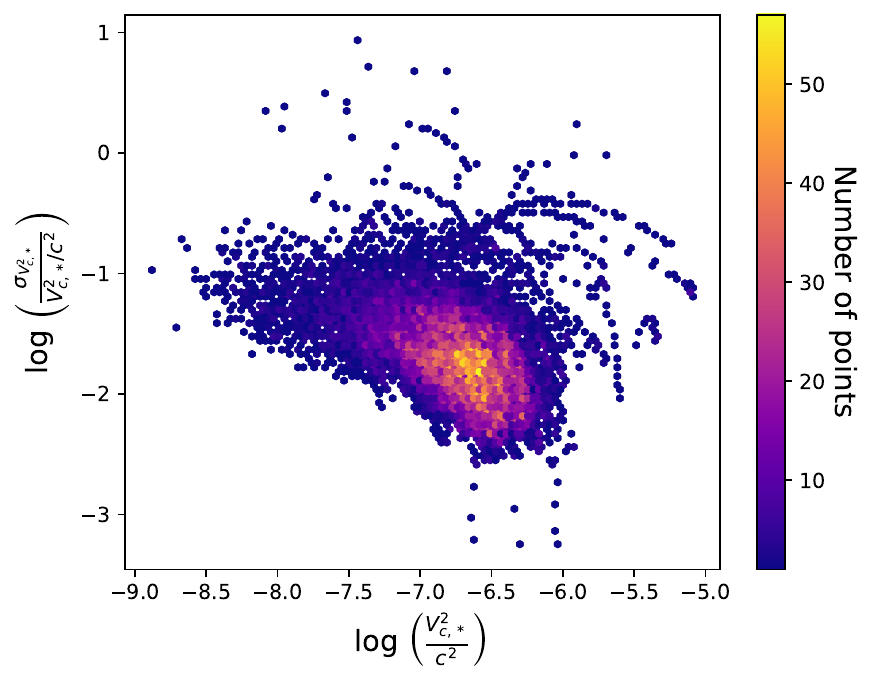}
    \caption{ Color map of intrinsic stellar circular velocity (divided by the speed of light), $V_{c,*}/c^2$,  and ratio of uncertainty of $V_{c,*}/c^2$ to its value $V_{c,*}/c^2$. The uncertainties are $\lesssim 10\%$ for most of the data points, indicating that variations of the mean values of the intrinsic circular velocity, $V_{c,*}/c^2$, used to classify the points in either screened or unscreened through Eq. (\ref{eq:chi}) do not affect the results (as also seen in Figure \ref{fig:hist_v2star}).
    }
    \label{fig:circ_err}
\end{figure}

\begin{figure}
    \centering
\includegraphics[width=\columnwidth]{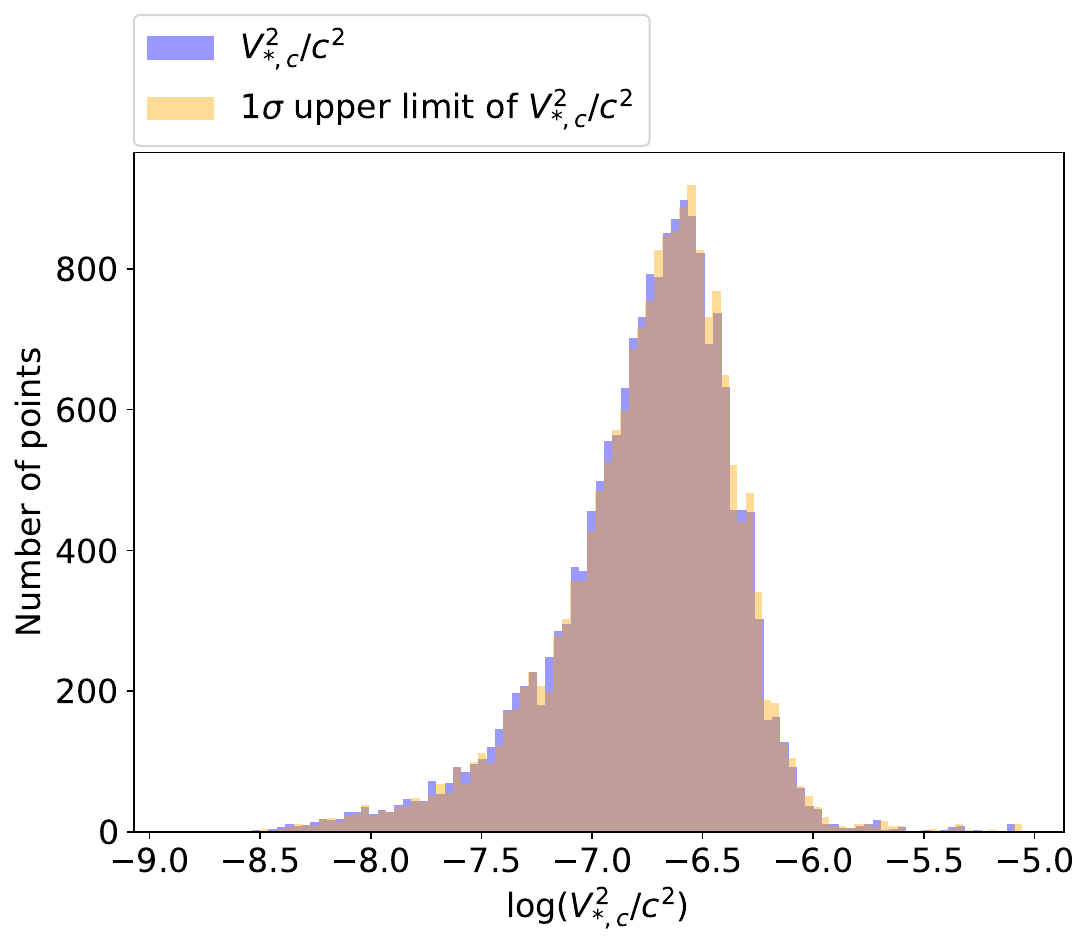}
    \caption{Intrinsic stellar circular velocity for all data points.}
    \label{fig:hist_v2star}
\end{figure}

We have also tested how the determination of the screening radius vary when using  mean plus one standard deviation on the intrinsic stellar circular velocity instead of the mean, and we find very similar results. The reason for this is because the uncertainties are relatively small for almost all data points for all galaxies: an example of the order of magnitude of the stellar circular velocity and its uncertainty is shown in Figure \ref{fig:circ_err}, while in Figure \ref{fig:hist_v2star} we show the histogram of the equivalent potential for the intrinsic stellar circular velocity and the $1\sigma$ upper limit of it. The 1$\sigma$ variation practically does not affect the distribution. Therefore, the uncertainty on the intrinsic circular velocity can be neglected when we  calculate the self-screening parameter. This determines uniquely the number of points in the screened and unscreened samples.    

Once we have the screened and unscreened samples for a given $\chi_c$, we constrain $\Delta G/G_N$ (along with other two parameters described below) using MCMC sampling.
We have two terms for the logarithm of the likelihood, one for the screened sample and other for the unscreened one. Theoretically $\Delta G/G_\text{N}$ should be positive, and nonzero only for unscreened points. The screened sample, in practice, does not necessarily have $V_{c,g}^2 = V_{c,*}^2$ due to intrinsic dynamical properties for the tracer, residual systematics (e.g. imperfectly modelled AD) or any other effect besides MG. In order to ensure that these effects do not bias our inference of $\Delta G/G_\text{N}$ from the unscreened sample, we introduce a free parameter $\delta$ which accounts for the difference between $V_{c,g}^2$ and $V_{c,*}^2$ on average over screened points. This parameter is learnt from the screened subsample but applied in the same way to the unscreened subsample so that the $\Delta G/G_\text{N}$ constraint comes entirely from the \emph{difference} between screened and unscreened points. Note that this gives the screened points a crucial role within our inference, as opposed to e.g. \citet{2020PhRvD.102j4060D} where they do not directly impact the $\Delta G/G_\text{N}$ constraint at all. Here, if the
number of points in the screened sample is small, the constraining power on $\delta$ and hence on $\Delta G/G_N$ will deteriorate.
$\delta$ also helps to calibrate the poorer beam-smearing correction for gas velocity curves at low radii.  

A second safeguard that we introduce into our likelihood is an additional scatter term $\sigma$ that we constrain simultaneously with the other parameters. This accommodates sources of uncertainty not explicitly accounted for in the {\tt NIRVANA} noise model, helping to prevent model misspecification leading to spuriously tight constraints on the other parameters.

Combining these approaches, our likelihood $\mathcal{L}$ is given by
\begin{align}\label{eq:likl}
\log(\mathcal{L}) &=- \sum_{{\rm unscr},i}\Bigg\{\frac{\big(\Delta G/G_\text{N} + 1+\delta - [V_{c,g}^2/V_{c,*}^2]_{ i}\big)^2}{2(\sigma_{{\rm err}, i}^2 + \sigma^2)}\nonumber\\
&+\frac{1}{2}\log(\sigma_{{\rm err}, i}^2 + \sigma^2) \Bigg\}\nonumber\\& -\sum_{{\rm scr},j}\Bigg\{\frac{\big(1+\delta - [V_{c,g}^2/V_{c,*}^2]_{ j}\big)^2}{2(\sigma_{{\rm err}, j}^2 + \sigma^2)}\nonumber\\
 & + \frac{1}{2}\log(\sigma_{{\rm err}, j}^2 + \sigma^2) \Bigg\}\,,
 \end{align}
where $i$  labels the data points of the unscreened sample, $j$ labels the data points of the screened sample and $\sigma_{\rm err}$ is the total uncertainty of one data point (which, as explained above, has a contribution from the estimated galaxy parameters and a contribution from the errors on the data themselves).
We use flat priors, with $0\leq \Delta G/G_N \leq 100$, $0\leq \sigma^2 \leq 100$ and $-100 \leq \delta \leq 100$. We run a MCMC using 20,000 steps and 10 walkers in the {\tt emcee} sampler \citep{emcee}. The convergence criterion is through the integrated autocorrelation time $\tau$, described in \citet{2010CAMCS...5...65G}, where chains longer than $\sim 50\tau$ are sufficient. The constraints are obtained through the {\tt corner.py} code \citep{corner}. 

The likelihood assumes that all data points are independent, even for the same galaxy. It is clear that when the points are from different galaxies there are no correlations between them, but for the same galaxy this would not be necessarily  true. Although {\tt NIRVANA} produces curves assuming that the points are independent,
we tested a simple scenario where the covariance matrix is non-diagonal with off-diagonal terms $0.8\sigma_i\sigma_j$, where $\sigma_i$ is the uncertainty previously calculated for the data point $i$, with $i\neq j$ and arbitrary constant $0.8$ multiplying the standard deviations. Although artificial, this covariance matrix enabled us to investigate how much our results would change if the data points were not independent. The constraints on $\Delta G/G_\text{N}$ for this off-diagonal scenario are slightly more stringent on average than the diagonal case.  Therefore, we can restrict our analysis to the most conservative situation, where all data points are independent of each other.

\begin{figure}
    \centering
    \includegraphics[width=\columnwidth]{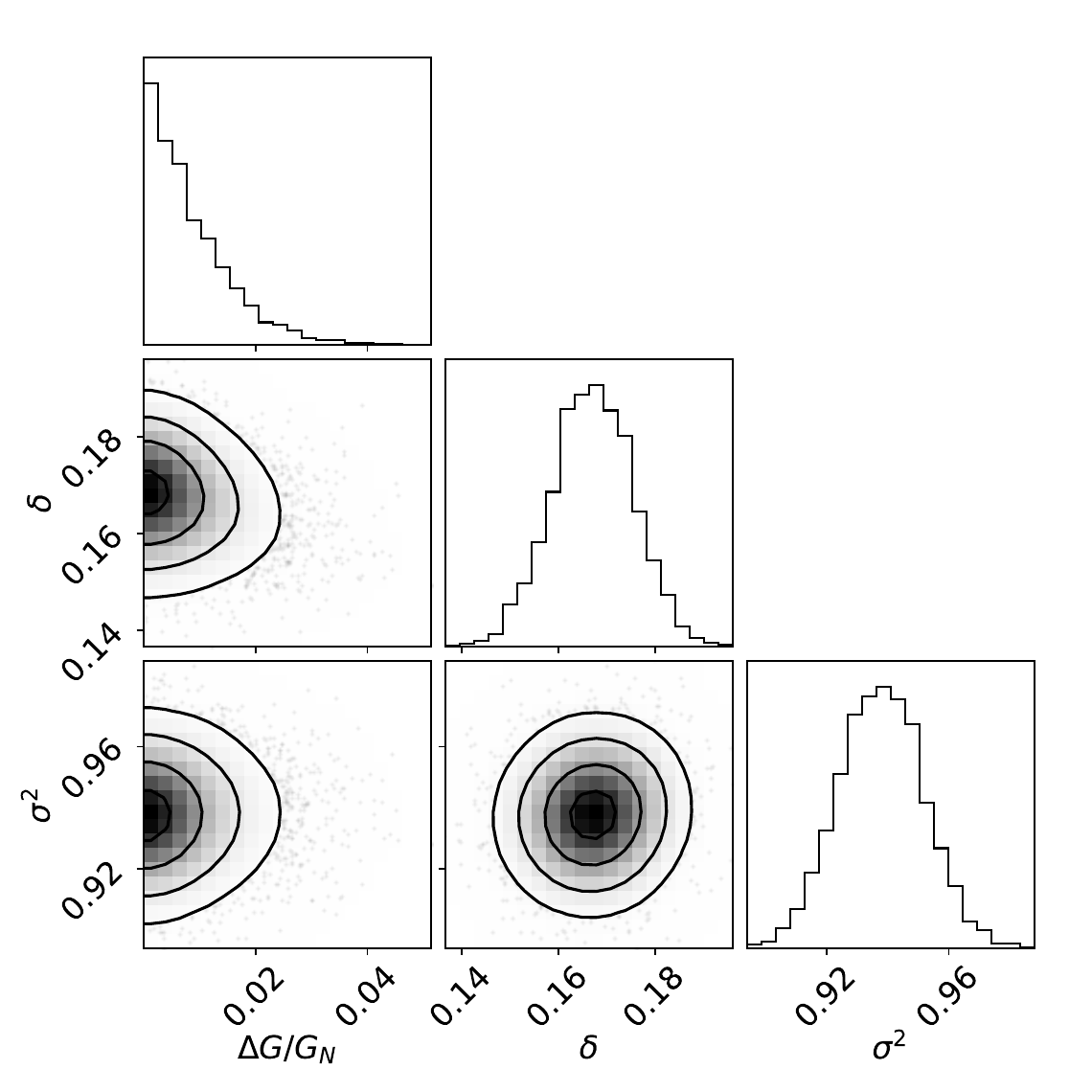}
    \caption{Example corner plot, for  $\chi_c=10^{-7}$.}
    \label{fig:corner}
\end{figure}

\section{Results}\label{sec:results}

\begin{figure}
    \centering
    \includegraphics[width=\columnwidth]{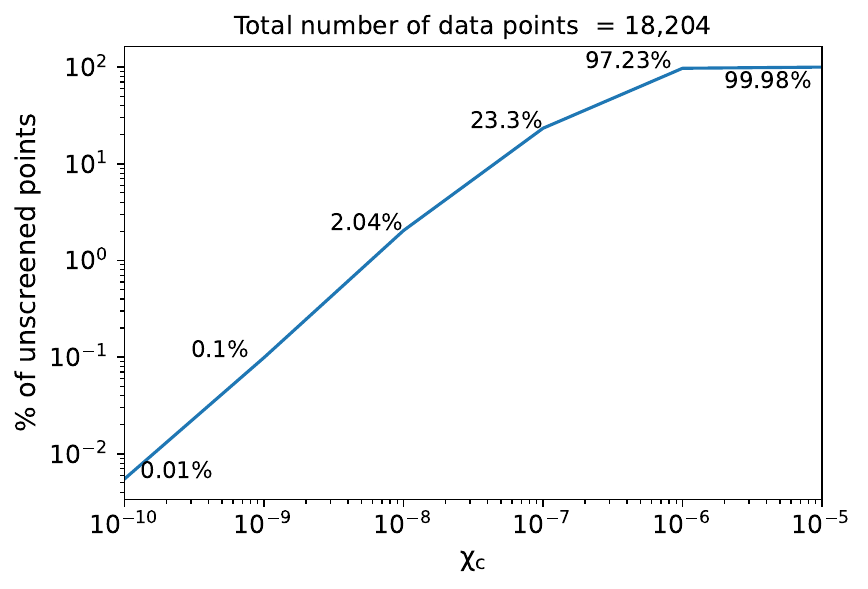}
    \caption{Fraction of unscreened points for different values of the self-screening parameter in the case without environmental screening.}
    \label{fig:n_unscr}
\end{figure}

\begin{figure}
    \centering
    \includegraphics[width=\columnwidth]{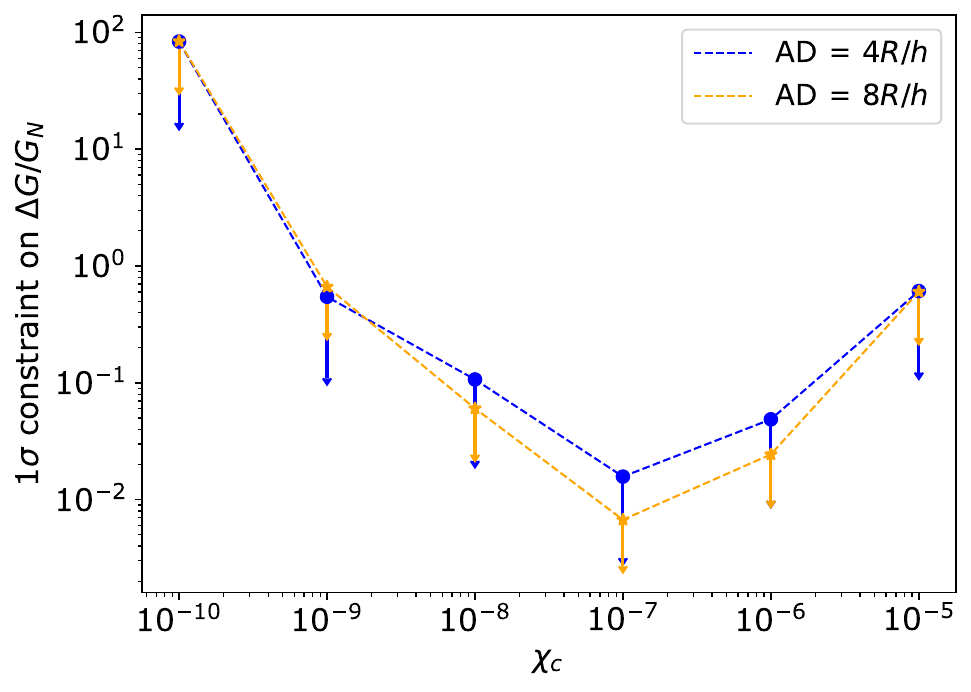}
    \caption{1$\sigma$ constraints on $\Delta G/G_\text{N}$, using the two different AD corrections for gas (Eq. \ref{eq:ad_final}), without environmental screening. The vertical arrows indicate that these are upper limits. 
    }
    \label{fig:ad}
\end{figure}

\begin{figure}
    \centering
    \includegraphics[width=\columnwidth]{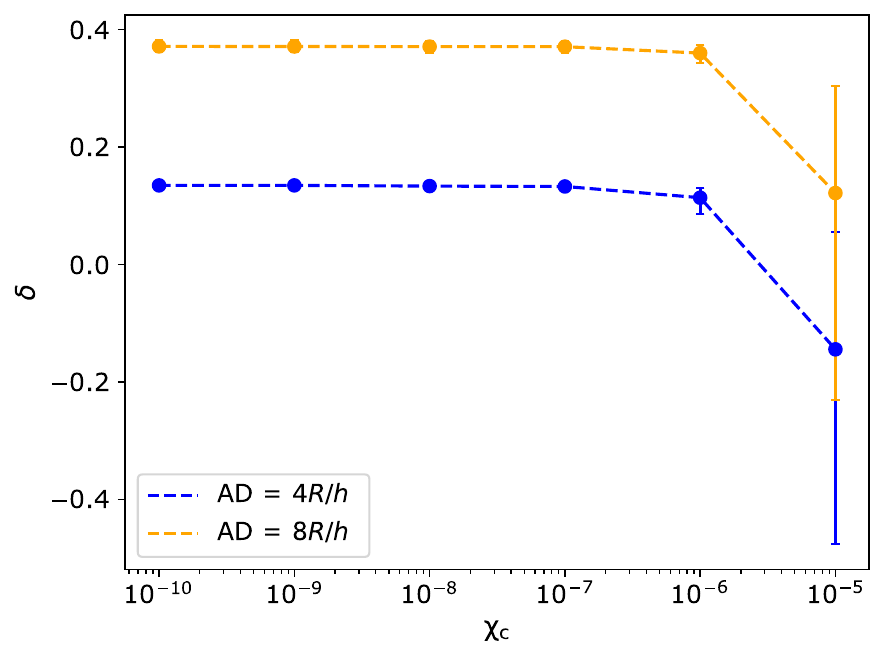}
    \caption{Constraints on $\delta$, using the two different AD corrections for gas (Eq. \ref{eq:ad_final}), without environmental screening. 1$\sigma$ limits are shown by the vertical error bars (very small for almost all $\chi_c$). }
    \label{fig:adsigmadelta}
\end{figure}

In order to investigate the two different AD corrections, we constrained the free parameters $\Delta G/G_\text{N}$, $\delta$ and $\sigma^2$,  for different values of $\chi_c$. An example corner plot with these constraints is shown in Figure \ref{fig:corner}. The fraction of points that are unscreened for each $\chi_c$ is shown in Figure \ref{fig:n_unscr}.  Beginning with the case without environmental screening for illustration, we show the results for $\Delta G/G_\text{N}$ and $\delta$ in Figures \ref{fig:ad} and \ref{fig:adsigmadelta}. The results for $\Delta G/G_\text{N}$ are very similar and the most conservative constraints come from AD correction with $4R/h$. For low values of $\chi_c$ most of the points are screened so the constraints on $\Delta G/G_\text{N}$ are very weak due to the lack of sufficient data. As the number of unscreened data points increases (larger $\chi_c$) the constraints become monotonically tighter reaching a point where both samples have a large amount of data points. After $\chi_c \sim 10^{-7}$ the number of screened points starts decreasing, therefore the constraints also deteriorate because $\delta$ is mainly constrained from the screened sample (but still influencing the results of the unscreened sample). The results for $\sigma^2$ and $\delta$ are roughly the same across different values of $\chi_c$, with the exception of when $\chi_c \approx 10^{-5}$ where the constraint on $\delta$ deteriorates due to paucity of screened points. The constraints on $\sigma^2$ are basically the same for all values of $\chi_c$ ($0.76\pm0.01$ for AD = $4R/h$ and $1.48\pm0.02$ for AD = $8R/h$).
Since both AD corrections give similar results and the one with an AD of $4R/h$ presents more conservative constraints on $\Delta G/G_\text{N}$, we will restrict ourselves from now on to this case for the gas circular velocity.

\begin{figure}
    \centering
    \includegraphics[width=\columnwidth]{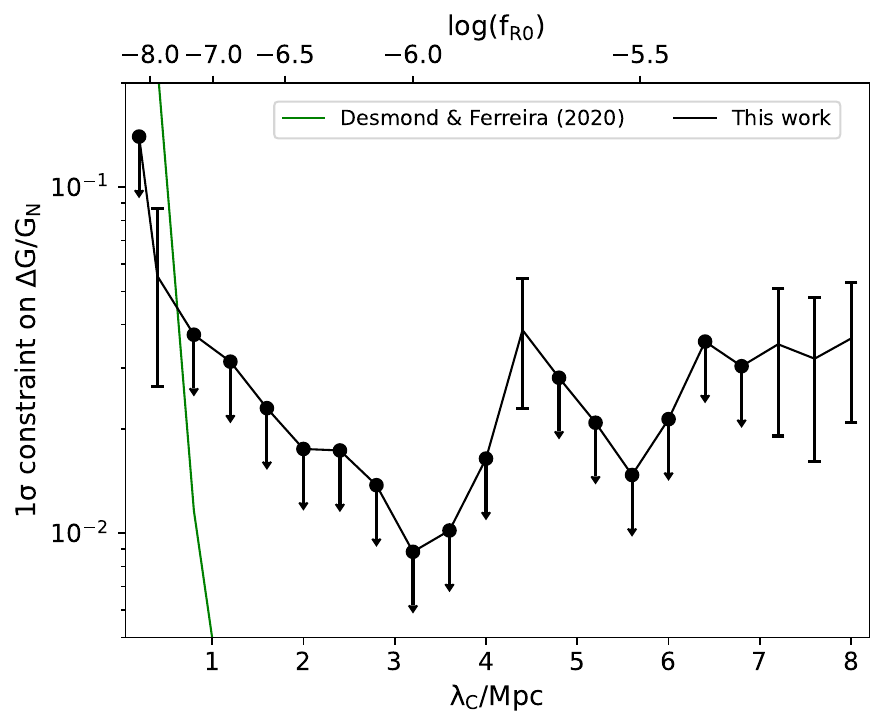}
    \caption{1$\sigma$ constraints on $\Delta G/G_\text{N}$ as a function of $\lambda_C$ or $f_{\mathcal{R}0}$ (black).  The green solid line is the combined constraints using gas--star offset and warp data obtained in \citet{2020PhRvD.102j4060D}, more powerful at higher values of $\lambda_C$ where a greater fraction of objects are unscreened. Vertical arrows denote 1$\sigma$ upper limits, while vertical bars represent $68\%$ C.L. upper and lower limits in case the $\Delta G/G_\text{N}$ posterior is formally discrepant with 0 at $>2\sigma$.}
    \label{fig:env1only}
\end{figure}

\begin{figure}
    \centering
    \includegraphics[width=\columnwidth]{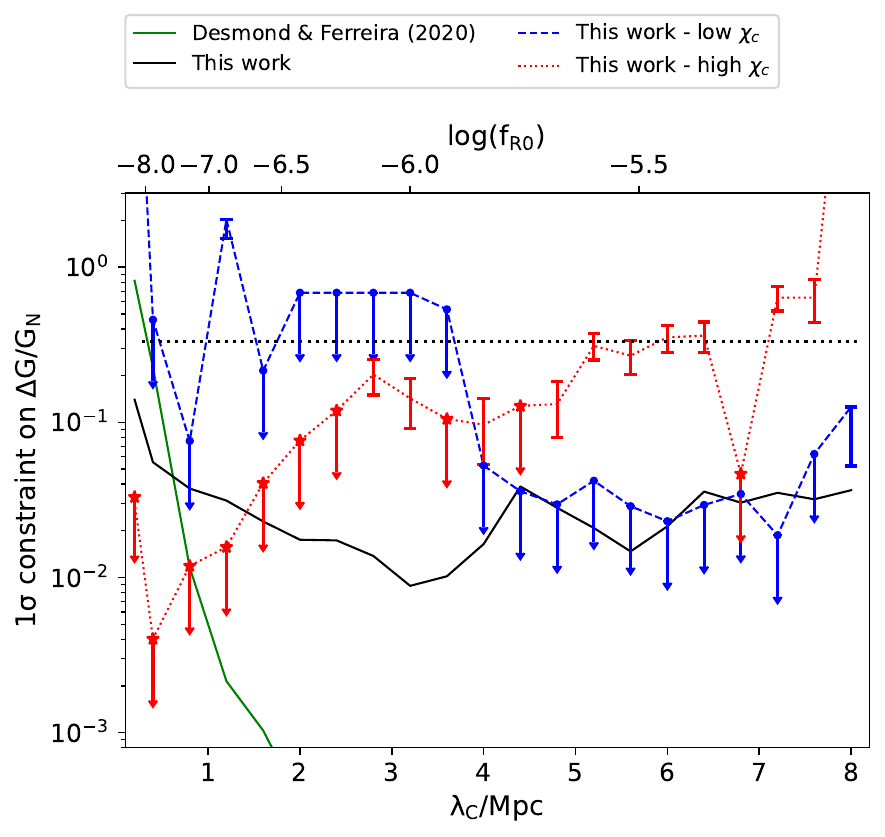}
    \caption{Similar to Figure \ref{fig:env1only}, but the red and blue lines show constraints  with $\chi_c$ raised or lowered by a factor of 10 from its fiducial dependence on $\chi_c$ (black solid line). The dotted line at $\Delta G/G_\text{N} = 1/3$ is the theoretical prediction of  the HS model.
     }
    \label{fig:env1}
\end{figure}

\begin{figure}
    \centering
    \includegraphics[width=\columnwidth]{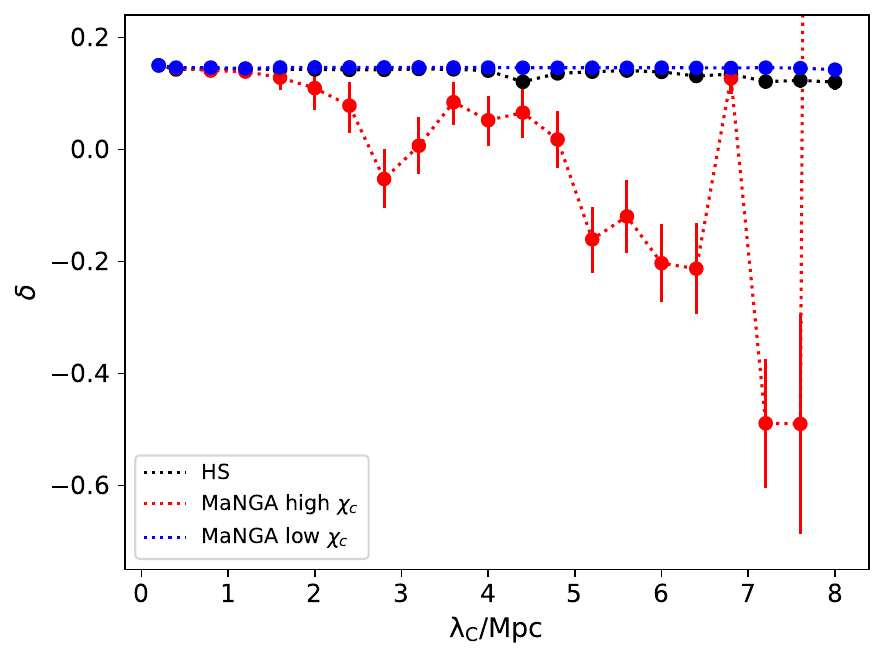}
    \caption{Constraints on $\delta$ as a function of $\lambda_C$ .    The red and blue lines show constraints   with $\chi_c$ raised or lowered by a factor of 10 from its fiducial dependence on $\chi_c$. The results are mostly overlapped and have small uncertainties. 
        }
    \label{fig:env1delta}
\end{figure}

\begin{figure}
    \centering
\includegraphics[width=\columnwidth]{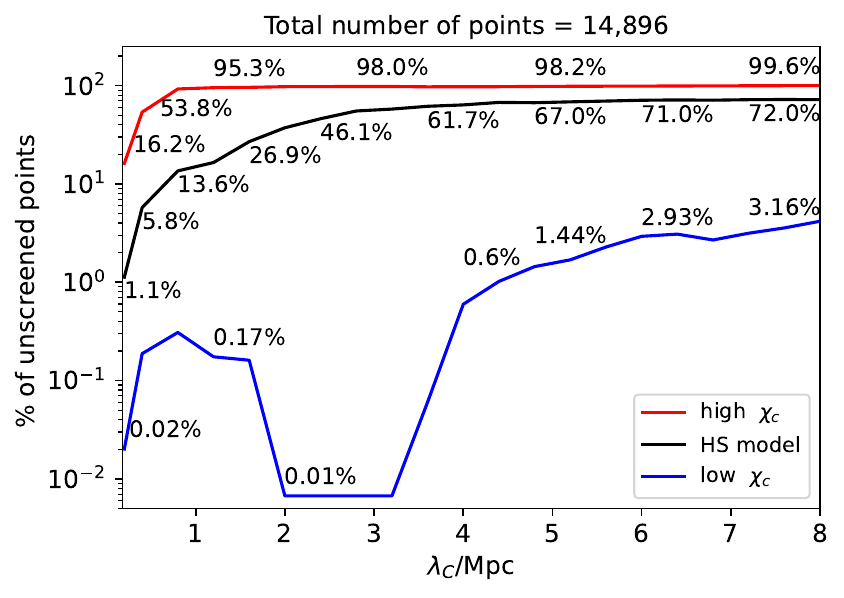}
    \caption{Fraction of unscreened points including environmental screening.  The red and blue lines show constraints  with $\chi_c$ raised or lowered by a factor of 10 from its fiducial dependence on $\chi_c$. 
    }
    \label{fig:n_unscr_env}
\end{figure}

When including environmental screening, we use as fiducial the HS model (Figure \ref{fig:env1only}), but also include a variation on the corresponding $\chi_c$ (thus departing from the fiducial model, in Figures \ref{fig:env1} and  \ref{fig:env1delta}). The fraction of unscreened points as a function of $\lambda_C$ is shown in Figure \ref{fig:n_unscr_env}. The total number of points is smaller than the only self-screening case because the environmental screening was calculated for galaxies with a distance from us of up to 250 Mpc (191 galaxies are removed from the sample, corresponding to 3308 points).

The results for $\Delta G/G_\text{N}$ present an oscillatory behaviour for some values of $\lambda_C$ and not the monotonic pattern seen before in the only self-screening scenario. These bumps are the results of small variations on the number of unscreened points in the sample when analysing the small variations on $\lambda_C$. The number of unscreened points is given by the screening radius evaluated from Eq. (\ref{eq:chi}), but with the left-hand side equals to $\chi_c -\Phi_{\rm ext}$. The standard behaviour without the environmental screening is that the screening radius  $r_s$ decreases as $\chi_c$ increases (and vice-versa), therefore the unscreened sample is larger for larger $\chi_c$ (in turn, larger $\lambda_C$). However, when the environmental screening is considered the self-screening parameter $\chi_c$ is no longer directly proportional to $r_s$, but it is related through the difference $\chi_c -\Phi_{\rm ext}$. Depending on the value of this difference the screening radius can vary in such a way that smaller values of $\chi_c$ do not necessarily increase the value of $r_s$. For instance, for the HS case,  there are 9,485 unscreened points (out of 14,896 points) for $\lambda_C=4$ Mpc , whereas for $\lambda_C=4.4 $ Mpc there are  10,054  unscreened points, and for $\lambda_C=4.8 $ Mpc there are 9,986 unscreened points. There are more unscreened points  for $\lambda_C=4.4 $ Mpc than for $\lambda_C=4.8$ Mpc (as opposed to the case without the environmental screening), because the difference $\chi_c -\Phi_{\rm ext}$ implies in a larger $r_s$ for $\lambda_C=4.8$ Mpc than for $\lambda_C=4.4$ Mpc. This causes a variation of the unscreened sample that influences the result in such a way that the unscreened sample may be smaller than the case without the environmental screening, causing the oscillatory behaviour seen in the results. This behaviour is what causes the unconstrained result for low $\chi_c$ around $\lambda_C= 3$ Mpc, because the unscreened sample has $\sim 20$ points for $\lambda_C\lesssim 2$ Mpc, and starts increasing from 6 unscreened points for  $\lambda_C\gtrsim 3$ Mpc. 

Finally, for the environmental screening case, there are values of $\Delta G/G_\text{N}$ that have a lower limit, depicted in the corresponding Figures by an error bar. The error bars are introduced when the 95\% C.L. are larger than zero. If the $2\sigma$ lower limit is negative we use an arrow to indicate $1 \sigma$ upper limit constraints. The mean values of $\Delta G/G_N$ and 68\% limits in the former case are shown in Table \ref{tab:detec}. Although there is a lower limit for some cases, they are few and most of them are still consistent with zero at 3 or 4$\sigma$. It is expected when testing a sufficiently large number of hypotheses that some of them will appear statistically significant even if not fundamentally so (the look-elsewhere effect), so we do not consider these ``detections'' significant. The constraints on $\sigma^2$ are roughly the same for all $\lambda_C$ ($0.66\pm 0.01$).

\begin{table}
    \centering
    \renewcommand{\arraystretch}{1.2}
    \begin{tabular}{c|c|c |c}
     \hline
           $\lambda_C$ (Mpc) & HS model  & $\lambda_C$ (Mpc) & Low $\chi_c$\\
           \hline
         0.4 & $0.055^{ +0.031 }_{ -0.029 }$ & 1.2 & $1.782^ { +0.249 }_{ -0.244 }$\\
        4.4 & $0.039^{ +0.016 }_{ -0.016 }$ & 8.0 & $0.089^ { +0.037 }_{ -0.036 }$\\
        7.2 & $0.035^{ +0.016 }_{ -0.016 }$ & & \\
        7.6 & $0.032^{ +0.016 }_{ -0.016 }$ & &\\
        8.0 & $0.037^{ +0.016 }_{ -0.016 }$  & &\\
         \hline
              $\lambda_C$ (Mpc) & High $\chi_c$  &    &  \\
         \hline
         2.8 & $0.203^ { +0.053 }_{ -0.053 }$ & 5.6 & $0.269^ { +0.066 }_{ -0.065 }$\\
    3.2 & $0.143^ { +0.050 }_{ -0.052 }$& 6.0 & $0.353^ { +0.070 }_{ -0.070 }$\\
    4.0 & $0.097^ { +0.045 }_{ -0.043 }$& 6.4 & $0.362^ { +0.081 }_{ -0.082 }$\\
    4.8 & $0.131^ { +0.052 }_{ -0.051 }$& 7.2 & $0.638^ { +0.116 }_{ -0.115 }$\\ 
    5.2 & $0.311^ { +0.061 }_{ -0.059 }$ &  7.6 & $0.637^ { +0.198 }_{ -0.195 }$ \\ 
    \hline 
    \end{tabular}
    \caption{Mean and 68\% limits of $\Delta G/G_N$ for the cases where there is an apparent detection ($\Delta G/G_N>0$ at $2\sigma$).}
    \label{tab:detec}
\end{table}

In the case of high $\chi_c$, the constraints on $\delta$ are towards negative values for larger values of $\lambda_C$ (Figure \ref{fig:env1delta}). In these cases, the contribution for screened points comes mostly  from the external potential. Depending on the value of the difference $\chi_c - \Phi_{\rm ext}$ the selected screened points present a stellar circular velocity larger than the gas circular velocity, resulting in negative values of $\delta$. 
The negative values of $\delta$ for those Compton wavelengths cause an increase on $\Delta G/G_N$ and hence an apparent detection. This is why  $\Delta G/G_N$ is $\sim 1$ for those $\lambda_C$'s (note that $\delta$ increases towards 0.2 for $\lambda_C = 6.8$ Mpc, causing $\Delta G/G_N$ to drop, being consistent with zero at 1$\sigma$).
The same fluctuation is seen for the case where the number of unscreened points is very low (low $\chi_c$ in Figure \ref{fig:env1}). There are less than 0.2$\%$  unscreened points in this region and they cause an oscillation and even a preference for $\Delta G/G_N\sim 1$.

Our results without the environmental screening excludes at $1\sigma$ confidence $\Delta G/G_N\gtrsim 0.1$ for $\chi_c\gtrsim10^{-8}$. On certain scales, $\chi_c \sim 10^{-7}$, the constraints even reach  $\sim 0.01$. When environmental screening is included in the analysis the constraints do not change significantly, i.e. $\Delta G/G_N<0.1$ for all scales considered, again sometimes reaching 0.01. Our results complement the analysis of \citet{2020PhRvD.102j4060D}, excluding regions $\lambda_C<0.3$ Mpc at $1\sigma$ confidence (equivalent to $f_{\mathcal{R}0}>10^{-8}$). For scales larger than $0.3$ Mpc our results do not impose better constraints than \citet{2020PhRvD.102j4060D}, mainly because here our constraining power on $\Delta G/G_N$ derives from both the screened and unscreened samples, while in \citet{2020PhRvD.102j4060D} the more unscreened galaxies the better.
Our results, therefore, are consistent with and complementary to \citet{2020PhRvD.102j4060D}, excluding the HS model on all relevant astrophysical scales.  


Our constraints are more stringent than all previous constraints besides those of~\citet{2020PhRvD.102j4060D}. Using galaxy rotation curves directly affords the bound $f_{\mathcal{R}0} < 10^{-6.1}$~\citep{2018MNRAS.480.5211N, 2019MNRAS.489..771N}, significantly weaker than our relative star-vs-gas test due to its strong degeneracy with the dark matter distribution.
Galaxy rotation curves have also been used to probe dark energy in the form of a cosmological constant, but without achieving competitive constraints as yet~\citep{Benisty_2}.
Cosmological probes achieve similar or even weaker constraints: using data from cosmic microwave background (CMB), supernovae and baryon acoustic oscillations (BAO) gives $|f_{\mathcal{R}0}|<10^{-4.79}$~\citep{2015PhRvD..92d4009C}, combining CMB, BAO and the Sunyaev--Zeldovich effect in clusters gives $f_{\mathcal{R}0}<3.7\times 10^{-6}$~\citep{2014PhRvD..90j3512B}, and using the gravitational wave signal emitted by the binary neutron star merger event GW170817 gives $f_{\mathcal{R}0}<3\times 10^{-3}$~\citep{2019PhRvD..99d4056J}. Combining cosmological constraints with those from the central regions of galaxies can strengthen them somewhat~\citep{Benisty_1}.

There are several systematic uncertainties in our analysis: the functional form of the velocity curves,  AD corrections, thin-disk model for the visible matter, covariance between data points and modelling assumptions in the likelihood.
However, we investigated these thoroughly and present here the results that we find to be conservative.


\section{Conclusions}\label{sec:conclusions}

Modified gravity is an alternative to the standard $\Lambda$CDM paradigm, aiming among other things to explain the current phase of accelerated expansion of our Universe. A broad class of models that evade Solar System  constraints are the ones presenting some sort of screening mechanism, where a fifth force effectively increases Newton's constant in low-density regions.

In this paper we have  used the SDSS-IV-MaNGA
dataset to search for a systematic excess of gas circular velocity over stellar circular velocity---expected in thin-shell-screened theories in the partially screened regime---and use it to constrain fifth-force parameters. Our method may be summarised as follows: 
\begin{enumerate}
    \item We selected 2220 blue galaxies with $M_\star > 5 \times 10^9 M_\odot$ in the range $0.007 < z < 0.15$.
    \item We used the open-source code {\tt NIRVANA} to extract the gaseous and stellar rotation curves and the velocity dispersion curves from the corresponding 2-D velocity maps, including the propagated uncertainties from the five relevant galaxy parameters. 
    \item We model both internal screening using the dynamical data itself and environmental screening using the gravitational maps constructed in~\citet{Desmond_reconstructing, 2020PhRvD.102j4060D} to distinguish screened and unscreened galaxies for given self-screening parameter (Eq.~\ref{eq:chi}).
    \item For each galaxy with reduced $\chi^2$ between 0.95 and 1.05, we average the rotation and dispersion velocities of each tracer (gas and stars) in bins of $\Delta R/R_{\rm eff}$. We then combine the rotation and dispersion of each tracer into the circular velocity curve by modelling AD (Eq.~\ref{eq:ad_final}), and calculate the ratio (gas over stars) for each galaxy.
     \item We used this data and the corresponding propagated uncertainties in MCMC to constrain the fifth-force strength and range, including marginalisation over astrophysical noise and a systematic offset in $V_\text{gas}/V_\text{star}$ across the sample.
\end{enumerate}

We extensively investigate potential systematics, namely, two different sets of functional forms for the velocity curves, two types of AD correction, a possible covariance between the data points and different assumptions for the likelihood. Our fiducial (and conservative) analysis uses a hyperbolic function for the rotation curve with an exponential velocity dispersion and AD correction given by Eq. (\ref{eq:ad_final}). Additionally we investigate separately the effects of internal and environmental screening. Our novel likelihood, which calibrates the model on the screened sample to isolate fifth-force effects, will likely be of use in future similar studies.

We do not find significant evidence for the effect of a fifth force, but rather
place a 1$\sigma$ upper limit to $\Delta G/G_\text{N}$ of $\sim$0.01--0.1 for all astrophysical fifth-force ranges. Our results are consistent with and complement literature fifth-force constraints, implying stringent restrictions on scalar--tensor theories. For example, for the  Hu--Sawicki model of $f(\mathcal{R})$ gravity we require $f_{\mathcal{R}0} \lesssim 10^{-8}$, rendering the theory astrophysically uninteresting.

Future more precise measurements of the velocity curves and determination of the screening potential could
improve the constraints, strengthening the bounds on $\Delta G/G_N$. We look to future IFU surveys capable of probing both stars and gas to achieve this. 
It will not however be possible to put meaningful bounds on $\Delta G/G_N$ for smaller values of $\lambda_C$, $f_{\mathcal{R}0}$ or $\chi_c$ because at lower values even the outer regions of dwarf galaxies in voids will become screened.

\section*{Acknowledgements}

The authors are very grateful to Kyle B. Westfall for valuable discussions and explanations about asymmetric drift, {\tt NIRVANA} and the MaNGA data analysis. We are grateful to Tariq Yasin for important input regarding the MaNGA galaxies, and for comments on the draft. We acknowledge earlier work by Karen L. Masters on this topic (private communication), and are grateful for discussions with her during the early stages of the project. We also thank Clare Burrage, Martin Bureau and Jeremy Sakstein for useful feedback and discussions. RL and HD are supported by Royal Society University
Research Fellowship grant no. 211046.  KK is supported by STFC grant ST/W001225/1. For the purpose of open access, we have applied a Creative Commons Attribution (CC BY) licence to any
Author Accepted Manuscript version arising.

Funding for the Sloan Digital Sky Survey IV has been provided by the Alfred P. Sloan Foundation, the U.S. Department of Energy Office of Science, and the Participating Institutions. SDSS acknowledges support and resources from the Center for High-Performance Computing at the University of Utah. The SDSS web site is \url{www.sdss4.org}.
SDSS is managed by the Astrophysical Research Consortium for the Participating Institutions of the SDSS Collaboration including the Brazilian Participation Group, the Carnegie Institution for Science, Carnegie Mellon University, Center for Astrophysics | Harvard \& Smithsonian (CfA), the Chilean Participation Group, the French Participation Group, Instituto de Astrofísica de Canarias, The Johns Hopkins University, Kavli Institute for the Physics and Mathematics of the Universe (IPMU) / University of Tokyo, the Korean Participation Group, Lawrence Berkeley National Laboratory, Leibniz Institut für Astrophysik Potsdam (AIP), Max-Planck-Institut für Astronomie (MPIA Heidelberg), Max-Planck-Institut für Astrophysik (MPA Garching), Max-Planck-Institut für Extraterrestrische Physik (MPE), National Astronomical Observatories of China, New Mexico State University, New York University, University of Notre Dame, Observatório Nacional / MCTI, The Ohio State University, Pennsylvania State University, Shanghai Astronomical Observatory, United Kingdom Participation Group, Universidad Nacional Autónoma de México, University of Arizona, University of Colorado Boulder, University of Oxford, University of Portsmouth, University of Utah, University of Virginia, University of Washington, University of Wisconsin, Vanderbilt University, and Yale University.

\section*{Data Availability}

The MaNGA data is publicly available here: \url{https://data.sdss.org/sas/dr17/manga/spectro/}. {\tt NIRVANA} can be installed via \url{https://github.com/ricardoclandim/NIRVANA} and other code used here can be found in \url{https://github.com/ricardoclandim/Testing_MG_Manga}.




\bibliographystyle{mnras}
\bibliography{mnras_main} 




 \appendix

 \section{Alternative model for velocity curves}\label{sec:appendix}

 We rerun the  analysis with an alternative model for the rotation and dispersion curves, replacing Eqs.~(\ref{eq:vrot_tanh}) and~(\ref{eq:vdisp_exp}):
\begin{align}
    V_{\rm rot, m}(R)&= V_0(1 - e^{-R/R_0}) \left(1 + V_1 \frac{R}{R_0}\right)\,,\label{eq:vrot_poly}\\ \sigma_{R,m}(R)&=\sigma_0e^{-R/h}+b\,,\label{eq:vdisp_expbase}
\end{align}
where $V_0$, $V_1$, $\sigma_0$, $R_0$, $h$ and $b$ are constants, estimated from {\tt NIRVANA} along with the five galaxy parameters. The total sample contains 991 galaxies (total of 16,181 points).
In this section we label the fiducial model as Model 1 and this alternative model as Model 2.

\begin{figure}
    \centering
  \includegraphics[width=\columnwidth]{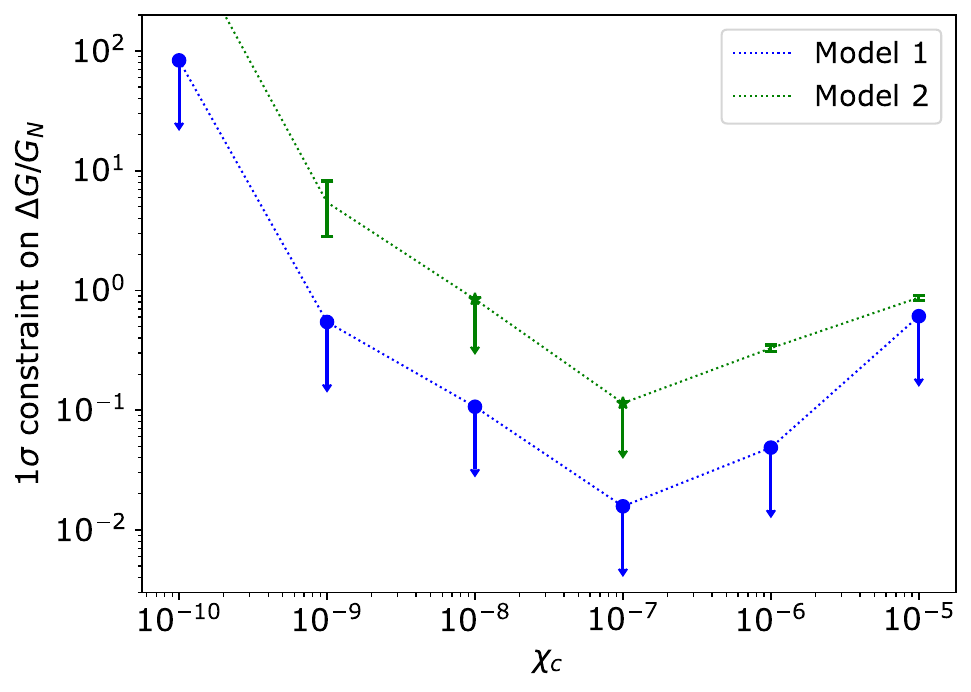}
    \caption{1$\sigma$ constraints on $\Delta G/G_\text{N}$ for Models 1 and 2, without environmental screening. The plot for Model 1 is the same as in Fig \ref{fig:adsigmadelta}. The vertical arrows reinforce the representation of the 1$\sigma$ upper limits, while vertical bars represent the ($68\%$ C.L.) upper and lower limits, when applicable.   
    }
    \label{fig:models}
\end{figure}

\begin{figure}
    \centering
  \includegraphics[width=\columnwidth]{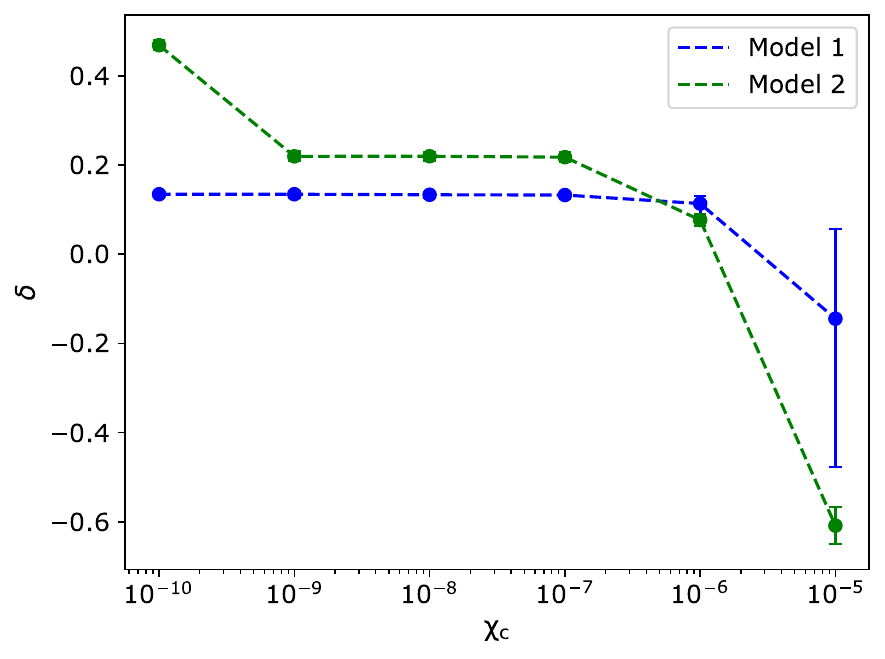}
    \caption{Constraints on $\delta$ for Models 1 and 2, without environmental screening. The plot for Model 1 is the same as in Fig \ref{fig:ad}. Vertical bars represent the upper and lower limits.  
    }
    \label{fig:modelsdeltasigma}
\end{figure}

We show in Figure \ref{fig:models} the 1$\sigma $ constraints on $\Delta G/G_\text{N}$ and in Figure \ref{fig:modelsdeltasigma} the constraints on $\delta$, as a function of the self-screening parameter, without environmental screening. The constraints on $\sigma^2$  for Model 2 are  $1.01\pm 0.02$ for all $\chi_c$. The samples for Model 2 are different from the ones of Model 1, because of the different modeling and therefore propagated uncertainties.  The weighted  average (where the weights are the inverse of the variances) for all points in Model 2 is $\langle V^2_{c,g}/V^2_{c, *}\rangle \approx 0.18$, whereas in Model 1 it is 0.69. The median in Model 1 is 1.06, whereas in Model 2 is 1.37. This indicates that the sample as a whole is roughly shifted, therefore constraints on $\delta$, for example, will give a more negative value than for Model 1. Model 1 gives tighter constraints than Model 2 for all  values of $\chi_c$. Model 2, on the other hand, presents positive lower limits on $\Delta G/G_\text{N}$ for some values of $\chi_c$.
More of the results for Model 2 indicate a detection of $\Delta G/G_\text{N}$ than for Model 1. However,
Model 1 fits the velocity curves  better than Model 2 (i.e. more galaxies have a reduced $\chi^2$ between 0.95 and 1.05). This is despite the fact that Model 1 has four free parameters while Model 2 has six. 

\begin{figure}
    \centering
    \includegraphics[width=\columnwidth]{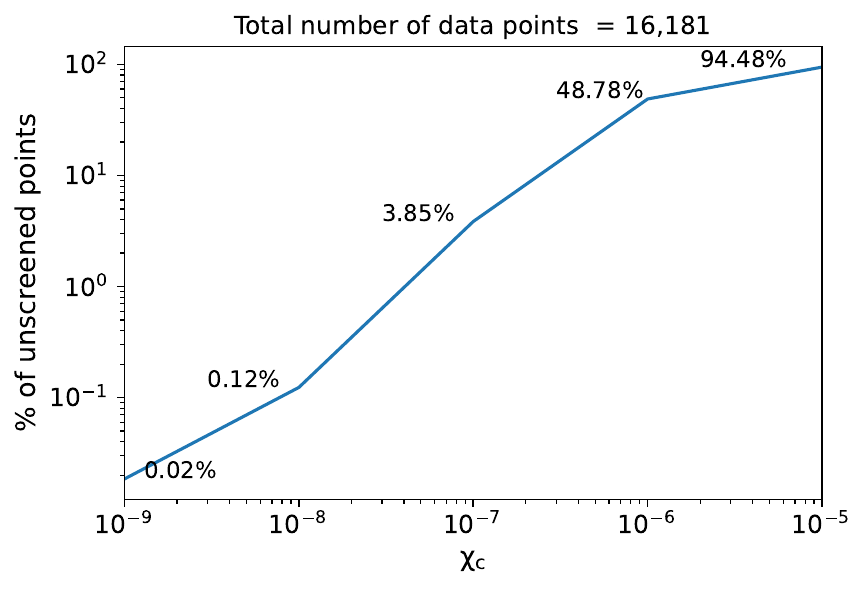}
    \caption{Same as Figure \ref{fig:n_unscr}, but for Model 2.}
    \label{fig:n_unscr2}
\end{figure}

\begin{figure}
    \centering
\includegraphics[width=\columnwidth]{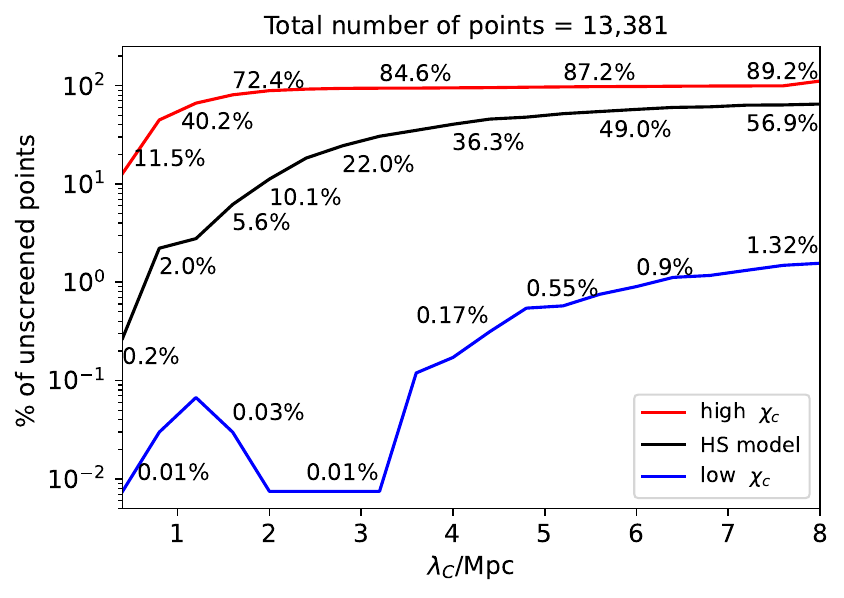}
    \caption{Same as Figure \ref{fig:n_unscr_env}, but for Model 2.}
    \label{fig:n_unscr_env2}
\end{figure}

The fraction of unscreened points for Model 2 with and without environmental screening is shown in Figures \ref{fig:n_unscr2} and \ref{fig:n_unscr_env2}, respectively. This is smaller than for Model 1.
In the case without environmental screening, only 0.02\% of all data points are unscreened for $\chi_c=10^{-5}$ for Model 1, whereas for Model 2 5.5\% are. This explains the small error bar in $\delta$, for this self-screening parameter, and the small error bars when environmental screening is included.

\begin{figure}
    \centering
    \includegraphics[width=\columnwidth]{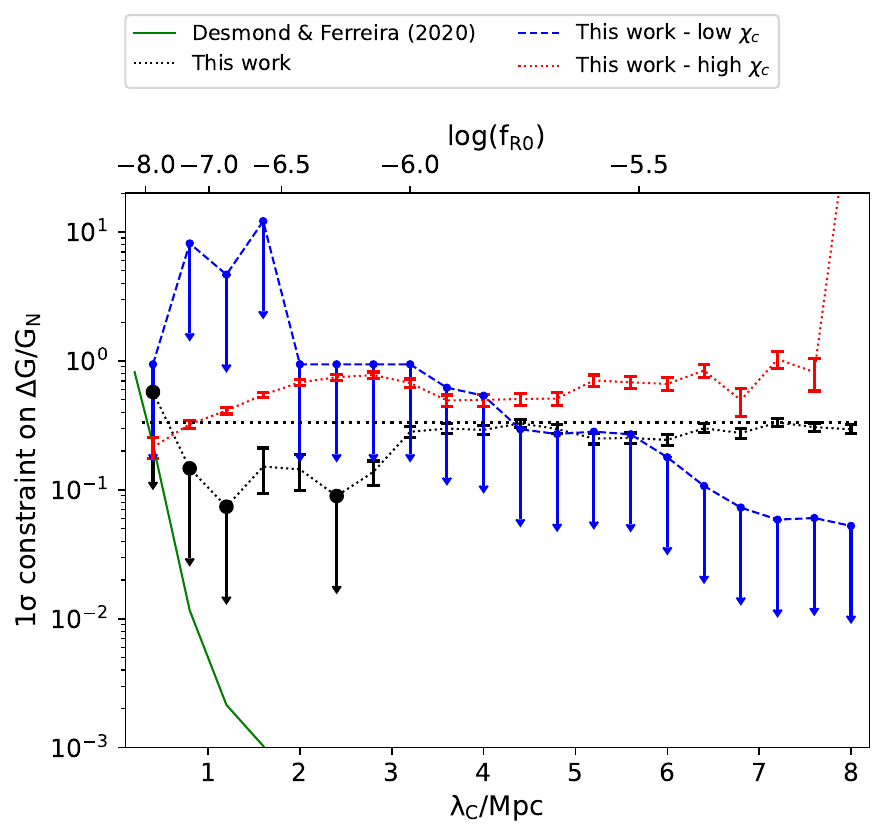}
    \caption{Same as Figure \ref{fig:env1}, but for Model 2.}
    \label{fig:env2}
\end{figure}

\begin{figure}
    \centering
    \includegraphics[width=\columnwidth]{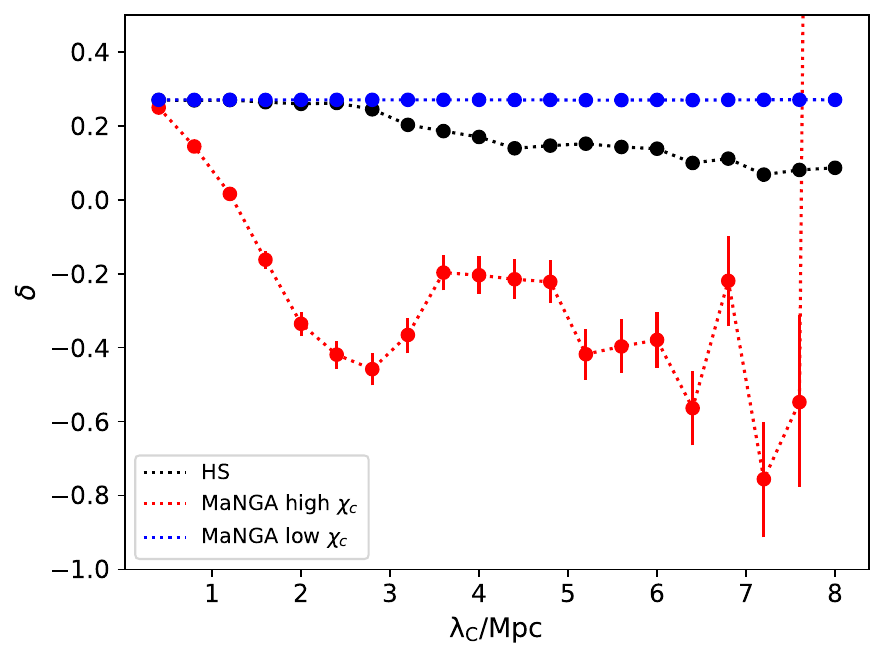}
    \caption{Same as Figure \ref{fig:env1delta}, but for Model 2. }
    \label{fig:env2delta}
\end{figure}

For the sake of completeness we show in Figures \ref{fig:env2} and \ref{fig:env2delta} the results when we include the environmental screening and in Table \ref{tab:detec2} values of $\Delta G/G_N$ when there is a detection. The constraints on $\sigma^2$ are $1.10\pm 0.02$ for all $\lambda_C$.

\begin{table}
    \centering
    \renewcommand{\arraystretch}{1.2}
    \begin{tabular}{c|c|c|c}
     \hline
     $\chi_c$ & Self-screening only & &\\
     \hline 
    $10^{-9}$ & $5.446^ { +2.757 }_{ -2.606 }$ & &\\
   $10^{-6}$ &$0.33^ { +0.021 }_{ -0.020 }$ & &\\
   $10^{-5}$ &$0.872^ { +0.042 }_{ -0.045 }$ & &\\
     \hline
           $\lambda_C$ (Mpc) & HS model \\
           \hline
        1.6 & $0.152^ { +0.06 }_{ -0.058 }$ & 5.2 & $0.25^ { +0.024 }_{ -0.023 }$\\
        2.0 & $0.144^ { +0.043 }_{ -0.045 }$& 5.6 & $0.252^ { +0.024 }_{ -0.023 }$\\
        2.8 & $0.138^ { +0.030 }_{ -0.030 }$&6.0 & $0.244^ { +0.024 }_{ -0.024 }$\\
        3.2 & $0.282^ { +0.028 }_{ -0.028 }$ & 6.4 & $0.301^ { +0.024 }_{ -0.024 }$\\
        3.6 & $0.299^ { +0.026 }_{ -0.026 }$ &6.8 & $0.276^ { +0.023 }_{ -0.024 }$\\
        4.0 & $0.292^ { +0.026 }_{ -0.025 }$& 7.2 & $0.333^ { +0.024 }_{ -0.024 }$\\ 
        4.4 & $0.328^ { +0.024 }_{ -0.024 }$& 7.6 & $0.306^ { +0.024 }_{ -0.024 }$\\
        4.8 & $0.296^ { +0.025 }_{ -0.024 }$&8.0 & $0.295^ { +0.024 }_{ -0.023 }$\\
         \hline
          $\lambda_C$ (Mpc) & High $\chi_c$ & &\\
         \hline
         0.4 & $0.214^ { +0.04 }_{ -0.039 }$ &  4.4 & $0.507^ { +0.054 }_{ -0.055 }$\\
        0.8 & $0.323^ { +0.024 }_{ -0.024 }$&4.8 & $0.511^ { +0.06 }_{ -0.06 }$\\
        1.2 & $0.412^ { +0.024 }_{ -0.024 }$ &5.2 & $0.706^ { +0.072 }_{ -0.068 }$\\
        1.6 & $0.546^ { +0.027 }_{ -0.028 }$& 5.6 & $0.683^ { +0.072 }_{ -0.075 }$\\
        2.0 & $0.682^ { +0.035 }_{ -0.035 }$&6.0 & $0.664^ { +0.078 }_{ -0.078 }$\\
        2.4 & $0.747^ { +0.040 }_{ -0.039 }$&6.4 & $0.843^ { +0.102 }_{ -0.101 }$\\
        2.8 & $0.777^ { +0.045 }_{ -0.046 }$& 6.8 & $0.495^ { +0.120 }_{ -0.123 }$\\
        3.2 & $0.674^ { +0.05 }_{ -0.047 }$& 7.2 & $1.031^ { +0.158 }_{ -0.155 }$\\
        3.6 & $0.494^ { +0.05 }_{ -0.050 }$&    7.6 & $0.82^ { +0.229 }_{ -0.236 }$\\
        4.0 & $0.498^ { +0.051 }_{ -0.053 }$& &\\ 
    \hline 
    \end{tabular}
    \caption{Same as Table \ref{tab:detec}, but for Model 2.}
    \label{tab:detec2}
\end{table}



\bsp	
\label{lastpage}
\end{document}